\documentclass[preprint, showpacs, prb]{revtex4}
\usepackage{hyperref}
\usepackage{amsmath}
\usepackage{amssymb}
\usepackage{graphicx}
\usepackage[nooneline]{subfigure}
\usepackage[cp1250]{inputenc}
\begin{document}
\title{A superconductor with 4-fermion attraction weakly perturbed by magnetic impurities}
\date{\today}
\author{Dawid Borycki}
\email{dawid.borycki@fizyka.umk.pl}
\affiliation{Instytut Fizyki, Uniwersytet M. Kopernika, ul. Grudziadzka 5, 87-100 Torun, Poland}

\begin{abstract}
A superconductor with 4-fermion attraction, considered by Ma\'{c}kowiak and Tarasewicz is modified by adding to the Hamiltonian a~long-range magnetic interaction $V$ between conduction fermions and localized distinguishable spin 1/2 magnetic impurities. $V$ has the form of a reduced s-d interaction. An upper and lower bound to the system's free energy density $f(H,\beta)$ is derived and the two bounds are shown to coalesce in the thermodynamic limit. The resulting mean-field equations for the gap $\Delta$ and a parameter $y$, characterizing the impurity subsystem are solved and the solution minimizing $f$ is found for various values of magnetic coupling constant $g$ and impurity concentration. The phase diagrams of the system are depicted with five distinct phases: the normal phase, unperturbed superconducting phase, perturbed superconducting phase with nonzero gap in the excitation spectrum, perturbed gapless superconducting phase and impurity phase with completely suppressed superconductivity.
\end{abstract}


\maketitle

\section{Introduction}
The properties of superconductors can be drastically changed by adding new elements, e.g. the superconducting transition temperature $T_{\text{c}}$ decreases in most superconductors, when magnetic impurities are added \cite{Matthias}. On the contrary, Bednorz and M\"{u}ller discovered that copper oxide doped with lanthanum and barium exhibits high $T_{\text{c}}\approx~35$~K \cite{BM}.

The effect of magnetic impurities on superconductors was studied by Abrikosov and Gor'kov \cite{AG}. Using Green's functions they not only explained the strong decrease in $T_{\text{c}}$ in the presence of magnetic impurities, but also predicted ''gapless behaviour'' of superconductors and complete destruction of superconductivity above critical concentration of impurities.

The problem of lowered $T_{\text{c}}$ in the presence of magnetic impurities was also discussed by other theorists, e.g. Nakamura \cite{Nakamura} and Suhl et al. \cite{Suhl} explained this effect by treating the s-d interaction $V_{\text{s-d}}$ \cite{Kasuya} as an additive term in the total Hamiltonian, which perturbes a BCS superconductor \cite{BCS}. Balseiro et al. \cite{Balseiro} studied a BCS superconductor perburbed by magnetic impurities interacting via a~nearest neighbour Heisenberg potential. The resulting phase diagrams comply qualitatively with experiment. Simon and others \cite{Simon} analyzed the disappearance of the Kondo effect in d-wave superconductors. Openov \cite{Openov} extensively discussed the question of reduced specific heat jump at $T_{\text{c}}$ in superconductors containing magnetic and nonmagnetic impurities. 

Our objective here is to study, the effect of magnetic impurities on a superconductor in a volume $|\Lambda|$, with 4-fermion attraction of the form
\begin{equation}
\label{V4f}
V_{\text{4f}} = -|\Lambda|^{-1} \sum_{\underline{k}\, \underline{k}'} G_{{\underline{k}\, \underline{k}'}}^{} b_{\underline{k}}^{*} b_{-\underline{k}}^{*} b_{-\underline{k}'}^{} b_{\underline{k}'}^{},
\end{equation}
where 
\[
b_{\underline{k}}^{} = a_{\underline{k}+}^{} a_{\underline{k}-}^{},
\]
and $a_{\underline{k}+}^{}$, $a_{\underline{k}-}^{}$ are fermion annihilation operators, whereas $G_{\underline{k}\, \underline{k}'}^{}$ is real, symmetric, invariant under the time reversal $\underline{k}\rightarrow - \underline{k}$ or $\underline{k}'\rightarrow - \underline{k}'$ and nonvanishing only in a thin band close to the Fermi surface, viz.,
\[
G_{\underline{k}\, \underline{k}'}^{} = G\chi(\underline{k})\chi(\underline{k}'), \qquad G > 0,
\]
where $\chi(\underline{k})$ denotes the characteristic function of the set
\[
S = \{\underline{k}: \mu - \delta \leq \varepsilon_{\underline{k}}^{} \leq \mu + \delta \}, \qquad \varepsilon_{\underline{k}}^{} = \frac{\hbar^2 k^2}{2m}.
\]

A superconductor with an interaction of this type was studied by Ma\'{c}kowiak et al. \cite{MT1, MT2, MT3, TM}. Tarasewicz et al. \cite{TB} demonstrated that 4-fermion attraction appears in sixth order expansion terms of Fr\"{o}hlich's transformation \cite{Frohlich}. So far, there is not much experimental evidence, which could confirm existence of fermion quartets generated by $V_{4f}$ in superconductors. An exception is the discovery of half-$h/2e$ magnetic flux quanta in SQUIDs \cite{Schneider}. 
Flux quanta with values close to $h \simeq 6/e$ and $h \simeq 8/e$ have also been observed. Accordingly, we can expect electrons to bind into quartets, sextet's, octets, and so on.

In this preliminary step of the investigation of a superconductor with 4-fermion attraction perturbed by magnetic impurities, we assume the perturbation implemented by the localized distinguishable spin $1/2$ magnetic impurities to be a reduced long-range s-d interaction, which involves only the z-components of the impurity and fermion spin operators. The reason for this simplification is that the thermodynamics of the resulting Hamiltonian $H=H_{0} + V_{4f} + V$ admits a mean-field solution which improves with decreasing impurity density. Furthermore, this solution is thermodynamically equivalent to the one obtained for $H$ with a Heisenberg type reduced s-d interaction
\begin{equation}
V_{H} = -\frac{g^2}{N} \sum_{\underline{k}\alpha} \left[(a_{\underline{k}-}^{*} a_{\underline{k}-}^{} - a_{\underline{k}+}^{*} a_{\underline{k}+}^{}) \sigma_{\alpha z}^{} - a_{\underline{k}+}^{*} a_{\underline{k}-}^{}\sigma_{\alpha -}^{} - a_{\underline{k}-}^{*} a_{\underline{k}+}^{}\sigma_{\alpha +}^{}\right], 
\end{equation}
\[
\sigma_{\alpha \pm}^{} = \sigma_{\alpha x}^{} \pm \text{i} \sigma_{\alpha y}^{},
\]
replacing $V$ (Ref.~\onlinecite{Mackowiak}, Sec. 6.2.5). (Similarly, the thermodynamics of classical superconductor can be explained in terms of a reduced BCS interaction, whereas a gauge-invariant theory of the Meissner effect requires a more general pairing potential.) The reduced form of $V_{H}$, obtained by rejecting the sum $\sum\limits_{\underline{k} \neq \underline{k}^{\prime}} V_{\underline{k} \, \underline{k}^{\prime}}$ in the s-d interaction $V_{\text{s-d}}$, is an approximation which resorts to the fact that spin-exchange processes, and not momentum exchange processes, are primarily responsible for the Kondo effect caused by $V_{\text{s-d}}$ \cite{Kondo}.

The interaction $V$ has the form
\begin{equation}
\label{Vred}
V = -\frac{g^2}{N} \sum _{\underline{k} \alpha} ( n_{\underline{k}-}^{} - n_{\underline{k}+}^{}) \sigma _{\alpha}^{}, \quad \sigma _{\alpha}^{} = \left( \begin{array} {cc} 1&0 \\ 0& -1 \\ \end{array} \right),
\end{equation}
$N$ denoting the number of magnetic impurities.
The interaction which induces the superconducting transition was chosen to be $V_{4f}$. The resulting theory is relatively simple and allows to construct the system's phase diagrams.
To this end, the system's free energy density $f(H,\beta)$ is bounded from above and below by mean-field type bounds, which are shown to be almost equal if the impurity density $d$ and width of conduction band $S$ are sufficiently small. The proof exploits methods developed by Czerwonko \cite{Czerwonko} and Tindemans et al. \cite{TC1, TC2}. The resulting mean-field equations, for the gap $\Delta$ and a parameter~$y$ characterizing the impurity subsystem, are solved and the solution, which minimizes $f$ is found for various values of magnetic coupling constant $g$ and density $d$. These equations have different forms in the limit as temperature tends to $0$K, which depend on $g$. For sufficiently small values of $g$, which preserve the weak perturbation by magnetic impurities, the limiting form of the gap equation, as $T \rightarrow 0$, is the same as in the BCS theory. In the present paper the weakness of reduced s-d coupling is assumed.

Phase diagrams of the system are depicted with five distinct phases: the normal phase, unperturbed superconducting phase, perturbed superconducting phase with nonzero gap in the excitation spectrum, perturbed gapless superconducting phase and impurity phase with completely suppressed superconductivity. 
An application of the theory to a superconductor with the electron-electron attraction of the form $V_{\text{BCS}} + V_{4f},$ is under construction.
\section{The Hamiltonian and lower bound to the free energy}
The full Hamiltonian of the system is
\begin{equation}
\label{Hamiltonian}
H = H_{s}^{\prime} + V, \qquad H_{s}^{\prime} = H_{0} + V_{\text{4f}},
\end{equation}
where $H^{}_{0} = \sum_{\underline{k} \sigma}\xi^{}_{\underline{k}} n^{}_{\underline{k}\sigma}$, with $\xi^{}_{\underline{k}} = \varepsilon^{}_{\underline{k}} - \mu$, $n^{}_{\underline{k}\sigma} = a^{*}_{\underline{k}\sigma}a^{}_{\underline{k}\sigma}$, is the kinetic energy of free fermions. 

\noindent In order to evaluate the free energy of the system
\[
{\cal{F}} = - \beta^{-1}\text{ln\,Tr\,exp}[-\beta H],
\]
\noindent
it is convienient to separate the fermion and impurity operators. This can be done by exploiting the identity
\[\begin{split}
\sum_{\underline{k}\alpha}\bigl(n_{\underline{k}-}^{}-n_{\underline{k}+}^{}\bigr)\sigma^{}_{\alpha} &=   \tfrac{1}{2}\sum_{\underline{k}\alpha}\bigl(n_{\underline{k}-}^{}-n_{\underline{k}+}^{}+\sigma^{}_{\alpha}\bigr)^2  \\ &- \tfrac{1}{2} \sum_{\underline{k}\alpha}(n_{\underline{k}-}^{} - 2n_{\underline{k}-}^{}n_{\underline{k}+}^{} + n_{\underline{k}+}^{})\\ &- \tfrac{1}{2}N|S|I,
\end{split}
\]
\noindent 
where $I$ stands for the operator
\[
I=I_1\otimes\dots\otimes{I_N},
\]
\noindent and $|S|$ denotes the number of points in $S$.
\noindent Let us now define
\[
\begin{split}
H_{s} &= H_{0} + V_{\text{4f}} + \tfrac{1}{2}g^{2}N^{-1} \sum_{\underline{k}\alpha}\bigl(n_{\underline{k}-}^{}- 2n_{\underline{k}-}^{}n_{\underline{k}+}^{}+n_{\underline{k}+}^{}\bigr) \\ &= H^{\prime}_{0} + \tfrac{1}{2}g^{2}|S|I,
\end{split}
\]
\noindent where
\begin{equation}
H_{0}^{\prime} = H^{}_{0} + V^{}_{4f} + \tfrac{1}{2}g^{2}N^{-1}\sum_{\underline{k}\alpha}\bigl(n_{\underline{k}-}^{}-
2n_{\underline{k}-}^{}n_{\underline{k}+}^{}+n_{\underline{k}+}^{}\bigr).
\end{equation}
Then $H$ can be written as
\begin{equation}
H = H_{s} + V^{\prime},
\end{equation}
where
\begin{equation}
\label{V}
V^{\prime}=-\frac{g^{2}}{2N}\sum_{\underline{k}\alpha}\bigl(n^{}_{\underline{k}-}-n^{}_{\underline{k}+}+\sigma^{}_{\alpha}\bigr)^{2}.
\end{equation}
To linearize the interaction (\ref{V}) we use the identity
\begin{equation}
\label{toz1}
\exp(a^{2})=\frac{1}{\sqrt{2\pi}}\int\limits_{-\infty}^{\infty}\exp\bigl(-\tfrac{1}{2}x^{2}+\sqrt{2}ax\bigr)\text{d}x.
\end{equation}
Then, exploiting the commutation relations
\begin{equation}
\bigl[H^{}_{s}, n_{\underline{k}-}^{} - n_{\underline{k}+}^{}\bigr] = \bigl[H^{}_{s}, \sigma^{}_{\alpha}\bigr] = 0,
\end{equation}
the statistical sum of the system
\begin{equation}
\begin{split}
\text{Z} &= \text{Tr}\exp(-\beta H) \\&= \text{Tr} \exp\Bigl[-\beta\Bigl(H^{}_{s}-\frac{g^{2}}{2N}\sum_{\underline{k}\alpha}
\bigl(n_{\underline{k}-}^{}-n_{\underline{k}+}^{}+\sigma^{}_{\alpha}\bigr)^{2}\Bigr)\Bigr],
\end{split}
\end{equation}
can be expressed as
\begin{widetext}
\begin{equation}
\label{Z}
\begin{split}
\text{Z} &= \, \text{Tr}\int\prod_{\underline{k}\alpha}\Bigl(\text{d}x^{}_{\underline{k}\alpha}\exp
\Bigl[-\tfrac{1}{2}N\beta x_{\underline{k}\alpha}^{2}+\beta g x^{}_{\underline{k}\alpha}
\bigl(n^{}_{\underline{k}}-n^{}_{\underline{k}+}+\sigma^{}_{\alpha}\bigr)\Bigr]\Bigl(\frac{N\beta}{2\pi}\Bigr)^{\frac{1}{2}}
\Bigr)\exp\bigl[-\beta H^{}_{s}\bigr] \\&= \,\Bigl(\frac{N\beta}{2\pi}\Bigr)^{\frac{N|S|}{2}}\int\prod_{\underline{k}\alpha}\text{d}x^{}_{\underline{k}\alpha}\exp\Bigl[-\tfrac{1}{2}\beta Nx_{\underline{k}\alpha}^{2} + G^{}_{\underline{k}\alpha}(x^{}_{\underline{k}\alpha})\Bigr],
\end{split}
\end{equation}
\end{widetext}
where $H_{s}=\sum_{\underline{k}}H_{s\underline{k}}=N^{-1}\sum_{\underline{k}\alpha}H_{s\underline{k}\alpha}$ and
\[
G^{}_{\underline{k}\alpha}(x^{}_{\underline{k}\alpha})= \ln\text{Tr}\exp\Bigl(-\beta N^{-1}H_{s\underline{k}\alpha}^{}+\beta g x_{\underline{k}\alpha}^{}\bigl(n_{\underline{k}-}^{}-
n_{\underline{k}+}^{}+\sigma_{\alpha}^{}\bigl)\Bigl).
\]
Following Pearce et al. \cite{Pearce}, let us now add and subtract $\frac{1}{2}N\beta\zeta^{-1}x_{\underline{k}\alpha}^{2}$, with $\zeta > 1$, in the exponent of the integrand on the r.h.s. of Eq. (\ref{Z}), and bound the resulting expression from above:
\begin{equation}
\label{ZBL}
\begin{split}
Z &\leq \prod_{\underline{k}\alpha} \max_{x_{\underline{k}\alpha}} \exp \left[ -\tfrac{1}{2} N \beta \zeta ^ {-1} x^{2}_{\underline{k}\alpha} + G^{}_{\underline{k}\alpha}(x^{}_{\underline{k} \alpha}) \right] \left(\frac{N\beta}{2\pi} \right)^{\frac{N|S|}{2}} \\&\times \int \prod_{\underline{k}^{\prime} \alpha^{\prime}} \exp \left[-\tfrac{1}{2} N \beta x^{2}_{\underline{k}^{\prime} \alpha^{\prime}}
\left(1 - \zeta^{-1} \right) \right] \text{d}x^{}_{\underline{k}^{\prime} \alpha^{\prime}}.
\end{split}
\end{equation}
Eq. (\ref{ZBL}) now yields a lower bound to the system's free energy $|\Lambda| f (H,\beta) = - \beta^{-1} \ln Z$:
\begin{equation}
\label{GestoscS}
|\Lambda |f(H,\beta)\geq \sum_{\underline{k}\alpha}\min_{x_{\underline{k}\alpha}}\Bigl(\tfrac{1}{2}\zeta^{-1}Nx_{\underline{k}\alpha}^{2}
-\beta^{-1}G^{}_{\underline{k}\alpha}\bigl(x^{}_{\underline{k}\alpha}\bigr)\Bigr)+\tfrac{1}{2}N|S|\beta^{-1}
\ln\bigl(1-\zeta^{-1}).
\end{equation}
The necessary condition for the minimum on the r.h.s. of Eq. (\ref{GestoscS}) is
\begin{equation}
\label{15}
\zeta^{-1}Nx^{}_{\underline{k}\alpha}=g\frac{\text{Tr}\bigl(n_{\underline{k}-}^{}-n_{\underline{k}+}^{}+\sigma^{}_{\alpha}\bigr)\exp\bigl[-\beta N^{-1}H^{}_{s\underline{k}\alpha}+\beta g x^{}_{\underline{k}\alpha}\bigl(n_{\underline{k}-}^{}-n_{\underline{k}+}^{}+\sigma^{}_{\alpha}\bigr)\bigr]}{\text{Tr}\exp\bigl[-\beta N^{-1}H^{}_{s\underline{k}\alpha}+\beta g x^{}_{\underline{k}\alpha}\bigl(n_{\underline{k}-}^{}-n_{\underline{k}+}^{}+\sigma^{}_{\alpha}\bigr)\bigr]}
\end{equation}
for $\underline{k} \in S, \alpha = 1, \dots, N$.
\section{Upper bound to the free energy}
An upper bound to the free energy of the system can be expressed in terms of the Hamiltonian $h$:
\begin{equation}
\label{male_h}
h = H_{s} + \tfrac{1}{2}N\sum_{\underline{k}\alpha}x^{2}_{\underline{k}\alpha} - g\sum_{\underline{k}\alpha}x^{}_{\underline{k}\alpha} 
\bigl(n_{\underline{k}-}^{}-n_{\underline{k}+}^{}+\sigma^{}_{\alpha}\bigr),
\end{equation}
where $x_{\underline{k}\alpha}\in\mathbb{R}$, is a solution of the equation
\begin{equation}
\label{Nx}
Nx^{}_{\underline{k}\alpha}=g\left<n_{\underline{k}-}^{}-n_{\underline{k}+}^{}+\sigma^{}_{\alpha}\right>_{h},
\end{equation}
with
\begin{equation}
\label{18}
\left<n_{\underline{k}-}^{}-n_{\underline{k}+}^{}+\sigma^{}_{\alpha}\right>_{h}=\frac{\text{Tr}\, \bigl(n_{\underline{k}-}^{}-n_{\underline{k}+}^{}+\sigma^{}_{\alpha}\bigr)\exp\bigl(-\beta h\bigr)}
{\text{Tr}\,\exp\bigl(-\beta h\bigr)}.
\end{equation}
The free energy of the system described by the Hamiltonian (\ref{male_h}) has the form
\begin{equation}
\label{gestosch}
\begin{split}
|\Lambda|f(h,\beta)&=\tfrac{1}{2}N\sum_{\underline{k}\alpha}x_{\underline{k}\alpha}^{2}-\beta^{-1}\ln\text{Tr}\exp\left[
-\beta H^{}_{s} + \beta g \sum_{\underline{k}\alpha}x^{}_{\underline{k}\alpha}\bigl(
n_{\underline{k}-}^{}-n_{\underline{k}+}^{}+\sigma^{}_{\alpha}\bigr)\right]\\
&=\tfrac{1}{2}N\sum_{\underline{k}\alpha}x_{\underline{k}\alpha}^{2}-\beta^{-1}\ln\prod_{\underline{k}\alpha}\text{Tr}\exp\Biggl[
-\beta N^{-1}H^{}_{s\underline{k}\alpha} + \beta g x^{}_{\underline{k}\alpha}\bigl(
n_{\underline{k}-}^{}-n_{\underline{k}+}^{}+\sigma^{}_{\alpha}\bigr)\Biggr]\\
&= \tfrac{1}{2}N\sum_{\underline{k}\alpha}x_{\underline{k}\alpha}^{2} - \beta^{-1}\sum_{\underline{k}\alpha}G^{}_{\underline{k}\alpha}\left(x^{}_{\underline{k}\alpha}\right).
\end{split}
\end{equation}
Bogolyubov's inequality
\begin{equation}
\label{20}
{\cal{F}}(H,\beta) \leq {\cal{F}}(h,\beta) + \left<H-h\right>_{h},
\end{equation}
yields the relevant upper bound.
Exploiting Eq. (\ref{Nx}) one obtains
\[
\left<H-h\right>_{h}=\left<V^{\prime}\right>_{h}-\tfrac{1}{2}N\sum_{\underline{k}\alpha}x_{\underline{k}\alpha}^{2}+N\sum_{\underline{k}\alpha}x_{\underline{k}\alpha}^{2} =\left<V^{\prime}\right>_{h}+\tfrac{1}{2}N\sum_{\underline{k}\alpha}x^2_{\underline{k}\alpha},
\]
where
\begin{equation}
\label{21}
\left<V^{\prime}\right>_{h}=-\tfrac{1}{2}g^{2}N^{-1} \sum_{\underline{k}\alpha}\left(\left<\left(n_{\underline{k}-}^{} - n_{\underline{k}+}^{}\right)^{2}\right>_{h}+2\left<\left(n_{\underline{k}-}^{} - n_{\underline{k}+}^{}\right)\sigma^{}_{\alpha}\right>_{h}+\left<\sigma^{2}_{\alpha}\right>_{h}\right).
\end{equation}
The inequality $\text{Tr}(\rho A^{2}) \geq \left(\text{Tr}(\rho A)\right)^{2}$, valid for any bounded self-adjoint operator $A$ and density matrix $\rho$, implies
\[
\left<V^{\prime}\right>_{h} \leq - \tfrac{1}{2} g^{2} N^{-1} \sum_{\underline{k} \alpha} \left< n^{}_{\underline{k}-} - n^{}_{\underline{k}+} + \sigma^{}_{\alpha} \right>^{2}_{h} = - \tfrac{1}{2} N \sum_{\underline{k}\alpha} x_{\underline{k} \alpha}^{2}.
\]
The thermal average $\left<H-h\right>_h$ is therefore bounded from above by $0$ and from Eq. (\ref{20}) we get
\begin{equation}
\label{22}
{\cal{F}}(H,\beta)\leq{\cal{F}}(h,\beta).
\end{equation}

\section{Comparison of the upper and lower bound to $f(H,\beta)$}
The form of Eqs. (\ref{15}), (\ref{Nx}) shows that their solutions $x_{\underline{k}\alpha}$ do not depend on $\underline{k}$, $\alpha$. Furthermore, for $|\zeta^{-1} - 1| \ll 1$ both equations have almost identical solutions. Henceforth, they will be denoted by $x$.
It follows now from the inequalities (\ref{GestoscS}), (\ref{22}) and Eq. (\ref{gestosch}) that the equality 
\begin{equation}
\label{f}
f(H,\beta) = \min_{x} f(h(x),\beta),
\end{equation}
where
\[
h(x) = H^{}_{s} + \tfrac{1}{2}N^{2}|S|x^{2} - gx\sum_{\underline{k}\alpha} \left(n^{}_{\underline{k}-} -  n^{}_{\underline{k}+} + \sigma^{}_{\alpha} \right),
\]
holds up to negligible terms if the density of magnetic impurities and the width $\delta$ of the conduction band $S$ are sufficiently small. On these grounds we shall assume that the thermodynamics of the original system is equivalent, under these restrictions, to that of $h(x)$ where $x$ is the minimizing solution of Eq. (\ref{Nx}).

The thermodynamics of a system with a Hamiltonian of the form $h(x)$ was studied in Refs.~\onlinecite{MT2, MT3, TM} . In the next section this method is exploited to evaluate $\lim \min \limits_{x} f(h(x),\beta)$ as $|\Lambda| \rightarrow \infty$.

\section{Thermodynamic equivalence of mean-field theory for $h(x)$}\label{sec5}
The thermodynamic perturbation method of Bogolyubov et al. \cite{B1, B2} , extended by Czerwonko in Ref.~\onlinecite{Czerwonko}, shows that in general \cite{MT3}
\begin{equation}
\label{lim}
\lim\limits_{|\Lambda| \rightarrow \infty} f(h(x),\beta) = \lim \limits_{|\Lambda| \rightarrow \infty} \min_{\{h_{0}\}} f(h_{0}(x),\beta),
\end{equation}
where
\begin{equation}
\label{h0}
h^{}_{0}(x) = \sum_{\underline{k} > 0} h^{}_{\underline{k}}(x) + h^{}_{i}(x) + C^{}_{0}
\end{equation}
and
\begin{equation}
\label{hk}
h_{\underline{k}}^{}(x)=\left(\xi_{\underline{k}}^{}+\tfrac{1}{2}g^{2}\right)\sum_{\sigma} \bigl( n_{\underline{k}\sigma}^{}+n_{-\underline{k}\sigma}^{}\bigr) - 2\Delta^{}_{\underline{k}} \bigl( \beta_{\underline{k}}^{} + \beta_{\underline{k}}^{*}\bigr) - g^{2}\bigl( n_{\underline{k}-}^{} n_{\underline{k}+}^{} + n_{-\underline{k}-}^{}n_{-\underline{k}+}^{}\bigr) + U_{\underline{k}}^{}(x) + C_{\underline{k}}^{},
\end{equation}
\begin{equation}
U_{\underline{k}}^{}(x)=-gxN\bigl(n_{\underline{k}-}^{}-n_{\underline{k}+}^{}+n_{-\underline{k}-}^{}
- n_{-\underline{k}+}^{}\bigl), \quad C_{\underline{k}}=2\Delta_{\underline{k}}\tau_{\underline{k}}, \quad C_{0}^{}=\tfrac{1}{2}|S|\bigl(g^{2} + N^{2}x^{2}\bigr),
\end{equation}
\begin{equation}
\label{Deltakk}
\Delta^{}_{\underline{k}} = |\Lambda| ^{-1} \sum_{\underline{k}^{\prime}} G_{\underline{k}\,\underline{k}^{\prime}}^{}\tau^{}_{\underline{k}^{\prime}}, \qquad h_{i}^{}(x) = \sum_{\alpha} h^{}_{\alpha}(x), \qquad h_{\alpha}^{}(x) = -gx|S|\sigma^{}_{\alpha},
\end{equation}
with $\xi^{}_{\underline{k}} = \varepsilon^{}_{\underline{k}} - \mu, \beta^{}_{\underline{k}} = b^{}_{-\underline{k}}b^{}_{\underline{k}}$.

The structure of $U_{\underline{k}}^{}(x)$ and $h_{\alpha}^{}(x)$ shows that $h^{}_{\underline{k}}(x)$ favours opposite alignment of impurity spins and those of conduction fermions. The interaction $V$ therefore acts as a quartet-breaker.

The difference
\[
\Delta h = h(x) - h_{0}^{}(x) = - |\Lambda|^{-1}\sum_{\underline{k}\,\underline{k}^{\prime}} G_{\underline{k}\,\underline{k}^{\prime}}^{}B_{\underline{k}}^{}B_{\underline{k}^{\prime}}^{*},
\]
where $B^{}_{\underline{k}} = \beta_{\underline{k}}^{} - \tau_{\underline{k}}^{}, \tau_{\underline{k}}^{}\in\mathbb{R}^{1}$, yields a negligible contribution to the limit on the r.h.s. of Eq.~(\ref{lim}), provided the constants $\tau_{\underline{k}}$ are adjusted to satisfy the condition
\begin{equation}
\label{Bk}
\left< B_{\underline{k}}^{}\right>_{h_{\underline{k}}}=\frac{\text{Tr} \left( B_{\underline{k}}^{}\exp \bigl[-\beta h_{\underline{k}}^{} (x)\bigr] \right)} {\text{Tr}\exp \bigl[-\beta h_{\underline{k}}^{}(x) \bigr]} = 0.
\end{equation}

\noindent According to Eqs. (\ref{lim}), (\ref{h0}) the system, can be equivalently described in terms of $h^{}_{0}(x)$ and to evaluate $f(h^{}_{0}(x),\beta)$ it suffices to diagonalize each $h^{}_{\underline{k}}$. This was done in Ref.~\onlinecite{MT3} by the method of Czerwonko~\cite{Czerwonko}. We briefly recapitulate the results.
The operator $h_{\underline{k}}^{}$ acts in the 16-dimensional space $M^{}_{\underline{k}}$ spanned by the vectors
\[
\left|n_{1}^{}n_{2}^{}n_{3}^{}n_{4}^{}\right> = \bigl(a_{\underline{k}+}^{*} \bigr)^{n_{1}} \bigl(a_{\underline{k}-}^{*} \bigr)^{n_{2}} \bigl(a_{-\underline{k}+}^{*} \bigr)^{n_{3}} \bigl(a_{-\underline{k}-}^{*} \bigr)^{n_{4}} \left|0\right>,
\]
where $n_{i}^{} = 0,1;\,i = 1,2,3,4$. Let us now define the spin operator
\begin{equation}
\label{SO}
2S^{}_{\underline{k}}=\sum_{\varphi=\pm 1}\sum_{\sigma=\pm 1} \sigma n_{\varphi\underline{k}\sigma}^{}
\end{equation}
and seniorities
\begin{equation}
\label{SE}
\Lambda_{\underline{k},\sigma}^{} = n_{\underline{k},\sigma}^{}-n_{-\underline{k},\sigma}^{}, \qquad \sigma = \pm,
\end{equation}
$S^{}_{\underline{k}}$ and $\Lambda^{}_{\underline{k},\sigma}$ commute with $h_{\underline{k}}^{}$:
\begin{equation}
\label{CR}
\left[h_{\underline{k}}^{}, 2S_{\underline{k}}^{}\right]=\left[h_{\underline{k}}^{},\Lambda_{\underline{k}\sigma}^{}\right] = 0.
\end{equation}
The commutation relations (\ref{CR}) enable diagonalization in the invariant subspaces of $M_{\underline{k}}$ with fixed eigenvalues of the operators $2S_{\underline{k}}^{}, \Lambda_{\underline{k}+}^{},\Lambda_{\underline{k}-}^{}$ and $h_{\underline{k}}^{}$. The space $M_{\underline{k}}$ splits into nine such invariant subspaces, viz.,
\begin{itemize}
\item[$-$]four 1-dimensional subspaces, which are spanned by the following eigenvectors with the corresponding eigenvalues $2s, \lambda_{\sigma}^{}, E_{\underline{k}}^{}$ of the operators $2S_{\underline{k}}^{}, \Lambda_{\underline{k},\sigma}^{}$ and $h_{\underline{k}}^{} - C_{\underline{k}}^{}$:

\[
\begin{array}{*{5}c}
\left|1010\right>\quad& 2s = 2\quad&\lambda_{+}^{}=1\quad&\lambda_{-}^{} = -1\quad&E_{\underline{k}}^{} = 2\xi_{\underline{k}}^{} + g^{2} + 2gNx\\\\
\left|0101\right> \quad& 2s = -2 \quad& \lambda_{+}^{} = -1 \quad& \lambda_{-}^{} = 1 \quad& E_{\underline{k}}^{} = 2\xi_{\underline{k}}^{} + g^{2} - 2gNx\\\\ \left|1100\right> \quad& 2s = 0 \quad& \lambda_{+}^{} = 1 \quad& \lambda_{-}^{} = 1 \quad& E_{\underline{k}}^{} = 2\xi_{\underline{k}}^{}\\\\ \left|0011\right> \quad& 2s = 0 \quad& \lambda_{+}^{} = -1 \quad& \lambda_{-}^{} = -1 \quad& E_{\underline{k}}^{} = 2\xi_{\underline{k}}^{}\\\\
\end{array}
\]
\item[$-$]four 2-dimensional subspaces, where the eigenvectors and eigenvalues are:
\[
\begin{array}{*{5}c}
\left|1000\right>\quad& 2s = 1\quad&\lambda_{+}^{}=1\quad&\lambda_{-}^{} = 0\quad&E_{\underline{k}}^{} = \xi_{\underline{k}}^{} + \frac{1}{2}g^{2} + gNx\\\\
\left|0001\right> \quad& 2s = -1 \quad& \lambda_{+}^{} = -1 \quad& \lambda_{-}^{} = 0 \quad& E_{\underline{k}}^{} = \xi_{\underline{k}}^{} + \frac{1}{2}g^{2} - gNx\\\\ \left|0010\right> \quad& 2s = 1 \quad& \lambda_{+}^{} = 0 \quad& \lambda_{-}^{} = -1 \quad& E_{\underline{k}}^{} = \xi_{\underline{k}}^{}+\frac{1}{2}g^{2}+gNx\\\\ \left|0100\right> \quad& 2s = -1 \quad& \lambda_{+}^{} = 0 \quad& \lambda_{-}^{} = 1 \quad& E_{\underline{k}}^{} = \xi_{\underline{k}}^{}+\frac{1}{2}g^{2}-gNx\\\\
\left|1110\right>\quad& 2s = 1\quad&\lambda_{+}^{}=1\quad&\lambda_{-}^{} = 0\quad&E_{\underline{k}}^{} = 3\xi_{\underline{k}}^{} + \frac{1}{2}g^{2} + gNx\\\\
\left|0111\right> \quad& 2s = -1 \quad& \lambda_{+}^{} = -1 \quad& \lambda_{-}^{} = 0 \quad& E_{\underline{k}}^{} = 3\xi_{\underline{k}}^{} + \frac{1}{2}g^{2} - gNx\\\\ \left|1011\right> \quad& 2s = 1 \quad& \lambda_{+}^{} = 0 \quad& \lambda_{-}^{} = -1 \quad& E_{\underline{k}}^{} = 3\xi_{\underline{k}}^{}+\frac{1}{2}g^{2}+gNx\\\\ \left|1101\right> \quad& 2s = -1 \quad& \lambda_{+}^{} = 0 \quad& \lambda_{-}^{} = 1 \quad& E_{\underline{k}}^{} = 3\xi_{\underline{k}}^{}+\frac{1}{2}g^{2}-gNx\\
\end{array}
\]
\item[$-$]one 4-dimensional subspace $M_{\underline{k}9}^{}$ spanned by the vectors
\[
\left|0000\right>\equiv\left|1\right>,\quad\left|1001\right>\equiv\left|2\right>,\quad \left|0110\right>\equiv\left|3\right>,\quad \left|1111\right>\equiv\left|4\right>,
\]
\end{itemize}
and where $2s=\lambda_{+}^{}=\lambda_{-}^{}=0$. If we choose for the basis of the $M_{\underline{k}9}^{}$ the following set of vectors: $\left\{\left|1\right>, -\left|2\right>,\left|3\right>, -\left|4\right>\right\}$ and denote the projector on $M_{\underline{k}9}^{}$ by $P_{\underline{k}9}^{}$, then the matrix form of the operator $h_{\underline{k}}^{} - C_{\underline{k}}^{}$ is
\begin{equation}
\label{Macierz}
P_{\underline{k}9}^{}\bigl(h_{\underline{k}}^{}-C_{\underline{k}}^{}\bigr)P_{\underline{k}9}^{} = \left(\begin{array}{*{4}c}0&0&0&2\Delta_{\underline{k}}^{}\\ 0&2\xi_{\underline{k}}^{}+g^{2}&0&0\\ 0&0&2\xi_{\underline{k}}^{}+g^{2}&0\\ 2\Delta_{\underline{k}}^{}&0&0&4\xi_{\underline{k}}^{}\end{array}\right).
\end{equation}
This leads to the secular equation
\begin{equation}
\bigl(2\xi_{\underline{k}}^{}+g^{2}-E\bigr)^{2}\bigl[E^{2} - 4\xi_{\underline{k}}^{}E - 4\Delta_{\underline{k}}^{2}\bigr]=0,
\end{equation}
which yields the following eigenvectors and eigenvalues of the operator (\ref{Macierz}):
\[
\begin{array}{cc}
\left|1001\right>\qquad&2\xi_{\underline{k}}^{}+g^{2}\\\\
\left|0110\right>\qquad&2\xi_{\underline{k}}^{}+g^{2}\\\\
u_{\underline{k}}^{} \left|0000\right> + v_{\underline{k}}^{} \left|1111\right> \qquad&2\xi_{\underline{k}}^{}-2E_{G\underline{k}}^{}\\\\
u_{\underline{k}}^{} \left|1111\right> - v_{\underline{k}}^{} \left|0000\right> \qquad&2\xi_{\underline{k}}^{}+2E_{G\underline{k}}^{}\\\\
\end{array}
\]
where
\[
u_{\underline{k}}^{2}=\frac{1}{2}\left(1+\frac{\xi_{\underline{k}}^{}}{E_{G\underline{k}}}\right), \qquad v_{\underline{k}}^{2}=\frac{1}{2}\left(1-\frac{\xi_{\underline{k}}^{}}{E_{G\underline{k}}}\right), \qquad E_{G\underline{k}}^{}=\sqrt{\xi_{\underline{k}}^{2}+\Delta_{\underline{k}}^{2}}.
\]

\section{The statistical sum and order parameters}
The operators $h_{\underline{k}}^{}$ and $\sigma_{\alpha}^{}$ act in different subspaces, so the statistical sum of the system equals
\begin{equation}
\label{StatSum}
\begin{split}
Z&=\prod_{\underline{k}>0}Z_{\underline{k}}^{}\prod_{\alpha}Z_{\alpha}^{}=\prod_{\underline{k}>0}\text{Tr} \, \exp\left[-\beta\left(h_{\underline{k}}^{} + 2|S|^{-1}C_{0}^{} \right)\right]\prod_{\alpha}\text{Tr} \exp\left[\beta gx|S|\sigma_{\alpha}^{}\right]\\ &=\exp\bigl[-\beta C_{0}^{}\bigr]\left[2\cosh\left(\beta gx|S|\right)\right]^{N}\prod_{\underline{k}>0}\text{Tr}\exp[-\beta h_{\underline{k}}^{}].
\end{split}
\end{equation}
The eigenstructure of $h_{\underline{k}}$, found in Sec. \ref{sec5}, yields
\begin{equation}\begin{split}
Z_{\underline{k}}^{} &= \text{Tr} \exp \left(-\beta h_{\underline{k}}^{} \right) = \exp \left[ -2\beta\left( \Delta_{\underline{k}}^{} \tau_{\underline{k}}^{} + \xi_{\underline{k}}^{} \right) \right] \Bigl[ 8 \exp \left(-\tfrac{1}{2} \beta g^{2} \right) \cosh(\beta \xi_{\underline{k}}^{}) \cosh(\beta g y) \\ &+  2 \exp \left(-\beta g^{2} \right) \left(1 + \cosh \left(2 \beta g y \right) \right) + 2 \cosh \left(2 \beta E_{G \underline{k}}^{} \right) + 2\Bigr],
\end{split}
\end{equation}
where $y = Nx$.
Eq. (\ref{Bk}) can be rewritten in the form
\begin{equation}
\label{tauk}
2 \tau^{}_{\underline{k}} = \beta^{-1} \frac{\partial}{\partial \Delta_{\underline{k}}}\text{ln} Z^{}_{\underline{k}}.
\end{equation}
Using Eqs. (\ref{Deltakk}), (\ref{tauk}) we get the equation for the order parameter
\begin{equation}
\label{Deltak}
\Delta_{\underline{k}}^{} = \tfrac{1}{2} |\Lambda|^{-1} \sum_{\underline{k}^{\prime}} G_{\underline{k} \,\underline{k}^{\prime}}^{} \frac{\Delta_{\underline{k}^{\prime}}}{E_{G\underline{k}^{\prime}}} F_{1}^{} \left( \beta, E_{G\underline{k}^{\prime}}, \xi_{\underline{k}^{\prime}}, y \right),
\end{equation}

\noindent where
\begin{equation}
\label{FunF}
F_{1}^{} \left(\beta, E_{G\underline{k}}^{}, \xi_{\underline{k}}^{}, y \right) = \frac{\sinh(2 \beta E_{G\underline{k}})}{4 \text{e}^{-\beta g^{2}/2} \cosh (\beta g y) \cosh (\beta \xi_{\underline{k}}) + \text{e}^{-\beta g^{2}} \left[1 + \cosh (2 \beta g y) \right] + \cosh (2 \beta E_{G\underline{k}}) + 1}.
\end{equation}
Passing from summation in Eq. (\ref{Deltak}) over $\underline{k}^{\prime}$ to integration over the single-fermion energies $\xi$ one obtains for a sufficiently thin band $S$:
\begin{equation}
\label{Delta}
\Delta^{} = \tfrac{1}{2} G \varrho \int \limits_{-\delta}^{\delta} \frac{\Delta}{E_{G}} F_{1}^{}(\beta, E_{G}^{}, \xi, y) \text{d}\xi,
\end{equation}
where $\varrho$ denotes the density of states in $S$. Equation (\ref{Delta}) is similar to the gap equation in BCS theory, however, the convexity properties of $F_{1}^{}(\beta, E_{G}^{}, \xi, y)$, with respect to $\Delta$, differ in general from those of $F_{\text{BCS}}^{}(\beta, E_{G}^{}) = \tanh (\beta E_{G}^{}/2)$ \cite{MT1}.

Given the spectrum of $h_{0}^{}$, one easily computes the thermal average
\begin{equation}
\left< n_{\underline{k}-}^{} - n_{\underline{k}+}^{} + \sigma_{\alpha}^{} \right>_{h_{0}}.
\end{equation}
From the commutation relations (\ref{CR}) one obtains
\begin{equation}
\left< n_{\underline{k}-}^{} - n_{\underline{k}+}^{} + \sigma_{\alpha}^{} \right>_{h_{0}} = \left< n_{\underline{k}-}^{} - n_{\underline{k}+}^{}\right>_{h_{\underline{k}}} + \left< \sigma_{\alpha}^{} \right> _{h_{\alpha}},
\end{equation}
where
\begin{equation}
\left< n_{\underline{k}-}^{} - n_{\underline{k}+}^{}\right>_{h_{\underline{k}}} = \frac{\text{Tr} \left( n_{\underline{k}-}^{} - n_{\underline{k}+}^{} + n_{-\underline{k}-}^{} - n_{-\underline{k}+}^{} \right) \exp [-\beta h_{\underline{k}}^{}]}{2 \, \text{Tr} \exp[-\beta h_{\underline{k}}^{}]}
\end{equation}
and therefore
\[
\frac{\text{Tr} \left( n_{\underline{k}-}^{} - n_{\underline{k}+}^{} + n_{-\underline{k}-}^{} - n_{-\underline{k}+}^{} \right) \exp [-\beta h_{\underline{k}}^{}]}{2 \, \text{Tr} \exp[-\beta h_{\underline{k}}^{}]} = \frac{(\beta g)^{-1} \frac{\partial}{\partial y} \text{Tr} \exp [-\beta h_{\underline{k}}^{}]}{2 \, \text{Tr} \exp [-\beta h_{\underline{k}}^{}]}.
\]
The eigenstructure of $h_{\underline{k}}^{}$ yields
\[ 
\begin{split}
\text{Tr} \exp [-\beta h_{\underline{k}}^{}] &= \exp \bigl[ -2 \beta (\Delta_{\underline{k}}^{} \tau_{\underline{k}}^{} + \xi_{\underline{k}}^{}) \bigr] \bigl[ 8 \exp (-\tfrac{1}{2} \beta g^{2}) \cosh (\beta \xi_{\underline{k}}^{}) \cosh(\beta g y) \\ &+ 2 \exp (-\beta g^{2}) \left( 1 + \cosh(2 \beta g y) \right) + 2 \cosh(2 \beta E_{G\underline{k}}^{}) + 2\bigr].
\end{split} 
\]
Hence 
\begin{equation}
\begin{split}
\left< n_{\underline{k}-}^{} - n_{\underline{k}+}^{} \right>_{h_{\underline{k}}} &= \bigl[2 \exp(-\tfrac{1}{2} \beta g^{2}) \cosh(\beta \xi_{\underline{k}}^{}) \sinh (\beta g y) + \exp(-\beta g^{2}) \sinh (2 \beta g y) \bigr]  \\ & \times \sinh ^{-1} (2 \beta E_{G\underline{k}}^{}) F(\beta, E_{G\underline{k}}^{}, \xi_{\underline{k}}^{}, y).
\end{split}
\end{equation}
Furthermore,
\[
\text{Tr} \exp [-\beta h_{\alpha}^{}] = 2 \cosh(\beta g y d^{-1}),
\]
\begin{equation}
\left< \sigma_{\alpha}^{} \right>_{h_{\alpha}}^{} = 2 \tanh (\beta g y d^{-1}).
\end{equation}
Therefore, Eq. (\ref{Nx}) takes the form
\begin{equation}
\label{y}
y = F_{2}(\beta, E_{G\underline{k}}^{}, \xi_{\underline{k}}^{}, y),
\end{equation}
where
\begin{equation}
\begin{split}
F_{2}(\beta, E_{G\underline{k}}^{}, \xi_{\underline{k}}^{}, y) &= g \frac{2 \text{e}^{- \beta g^{2}/2} \cosh(\beta \xi_{\underline{k}}^{}) \sinh(\beta g y) + \text{e}^{-\beta g^{2}} \sinh(2 \beta g y)}{4 \text{e}^{-\beta g^{2}/2} \cosh (\beta g y) \cosh (\beta \xi_{\underline{k}}^{}) + \text{e}^{-\beta g^{2}} \left[1 + \cosh (2 \beta g y) \right] + \cosh (2 \beta E_{G\underline{k}}^{}) + 1} \\ &+ g \tanh(\beta g y d^{-1}).
\end{split}
\end{equation}
Equations (\ref{Delta}) and (\ref{y}) constitute the set of two equations for $\Delta,\, y$. The expression for the statistical sum (\ref{StatSum}) and the equality
\[
\tau_{\underline{k}}^{} \Delta_{\underline{k}}^{} = \tfrac{1}{2} \Delta_{\underline{k}}^{2} E_{G\underline{k}}^{-1} F_{1}^{}(\beta, E_{G\underline{k}}^{}, \xi_{\underline{k}}^{}, y),
\]
resulting from Eq. (\ref{tauk}), as well as Eqs. (\ref{f}), (\ref{lim}), lead to the following expression for the free energy
\begin{equation}
\begin{split}
\cal{F} &= \min_{\{ \Delta_{\underline{k}}^{},\,y \}} \sum_{\underline{k} > 0} \Bigl[ \Delta^{2}_{\underline{k}} E_{G\underline{k}}^{-1} F_{1}^{}(\beta, E_{G\underline{k}}^{}, \xi_{\underline{k}}^{}, y) + y^{2} + g^{2} + 2\xi_{\underline{k}}^{} - \beta^{-1} \text{ln} P_{\underline{k}}^{} \Bigr] \\ & - 2 \sum_{\underline{k}>0} N (|S| \beta)^{-1} \text{ln} \bigl[2 \cosh(\beta g y d^{-1}) \bigr] \Bigr],
\end{split}
\end{equation}
where the minimum runs over all solutions of Eqs. (\ref{Delta}), (\ref{y}) and
\[
P_{\underline{k}}^{} = \Bigl[ 8 \text{e}^{-\beta g^{2}/2} \cosh (\beta \xi_{\underline{k}}^{}) \cosh(\beta g y) + 2 \text{e}^{-\beta g^{2}} \bigl(1 + \cosh(2 \beta g y) \bigr) + 2 \cosh(2\beta E_{G\underline{k}}^{}) + 2 \Bigr].
\]
If $G_{\underline{k}\,\underline{k}^{\prime}}^{}$ is nonvanishing and constant only in a thin band near the Fermi surface: $G_{\underline{k}\,\underline{k}^{\prime}}^{}=G\chi(\underline{k})\chi(\underline{k}^{\prime}),\,G>0$, then the free energy density equals
\begin{equation}
\label{Gestoscf}
\begin{split}
f(h,\beta) &= \min_{\Delta,\,y} \varrho \int \limits_{-\delta}^{\delta} \bigl[\tfrac{1}{2} \Delta^{2} E_{G}^{-1} F_{1}^{}(\beta, E_{G}^{}, \xi, y) + \tfrac{1}{2}(y^{2} + g^{2}) + \xi - \tfrac{1}{2} \beta^{-1} \text{ln} P \bigr] \text{d}\xi \\ & - \varrho d \beta^{-1} \int \limits_{-\delta}^{\delta} \text{ln} \bigl[2 \cosh(\beta g y d^{-1}) \bigr] \text{d}\xi + E_{0}^{}(\Delta = 0) + \varrho \delta^{2},
\end{split}
\end{equation}
where $d=N|S|^{-1}$ is proportional to impurity concentration and $E_{0}^{}(\Delta = 0)$ denotes the ground-state energy of free fermions. The last two terms are the contribution to the free energy density from one-fermion states, lying outside $S$.

Further sections contain numerical analysis of Eqs. (\ref{Delta}), (\ref{y}) and (\ref{Gestoscf}). The numerical solution of the set of Eqs. (\ref{Delta}), (\ref{y}) has been carried under the assumption that one of the two inequalities 
\[
-\frac{1}{2}g^{2} + gy + \xi < 2E_{G}
\]
or
\[
2 g y - g^{2} < 2E_{G}
\] 
holds at $T=0$K. Eq. (\ref{Deltak}) then takes the limiting form 
\begin{equation}
\label{Delta0}
G \varrho\, \text{arcsinh} \left(\frac{\delta}{\Delta(0)}\right) = 1
\end{equation}
at $T=0$K, whereas Eq. (\ref{y}) is satisfied by $y=g$ in this limit. The above inequalities state weak coupling of impurities to conducting fermions. 
Thus at sufficiently low temperatures $T$, close to $0$K, $\Delta(T)$ is taken as the solution of Eq. (\ref{Delta0}) and substituted into Eq.~(\ref{y}), which is then solved for one-fermion energies $\xi \in S$ by exploiting the Newton-Raphson method. The resulting values of $y(\xi)$ are used to obtain $\Delta(T+\Delta T)$ from Eq.~(\ref{Delta}) by deploying a Newton-Cotes quadrature as long as the result is self-consistent. The resulting value of $\Delta(T+\Delta T)$ is exploited to achieve $y(\xi)$ at temperature $T+\Delta T$. This procedure is continued until $T$ reaches the specified value.

The computations in Secs. \ref{SPD}, \ref{SCV} are performed under the assumption that $\mu = \varepsilon_{F}^{}$ and $\partial \mu / \partial T = 0$. Justification of this assumption is given in Sec. \ref{CP}. 
\section{Phase diagrams}
\label{SPD}
Equations (\ref{Delta}), (\ref{y}) clearly possess the solution $\Delta = y = 0$ at all values of $\beta \geq 0$. At sufficiently large values of $\beta$ one finds also other solutions, viz., $\{ \Delta \neq 0, y = 0 \}$, $\{ \Delta = 0, y \neq 0 \}$, $\{ \Delta \neq 0, y \neq 0 \}$. The system's state is characterized, according to Eq. (\ref{Gestoscf}), by the solution which minimizes $f(h,\beta)$. It will be denoted by $\{\Delta_{\text{m}}, y_{\text{m}}\}$. Accordingly, we distinguish the following phases:
\begin{itemize}
\item[$-$]{$\{\Delta_{\text{m}} = 0, y_{\text{m}} = 0 \}$ corresponds to the paramagnetic phase $P$}
\item[$-$]{$\{\Delta_{\text{m}} \neq 0, y_{\text{m}} = 0 \}$ corresponds to the unperturbed superconducting phase $S$}
\item[$-$]{$\{\Delta_{\text{m}} = 0, y_{\text{m}} \neq 0 \}$ describes the ferromagnetic phase $F$ without bound quartets, in which the impurity spins tend to align opposite to those of conduction fermions}
\item[$-$]{$\{\Delta_{\text{m}} \neq 0, y_{\text{m}} \neq 0 \}$ describes the $D$ phase in which superconductivity coexists with ferromagnetism}
\end{itemize}
We define the following temperatures corresponding to the respective phase transitions and characteristic points of $\Delta(T)$ plots:
\begin{itemize}
\item[$-$]{$T_{\text{c}}$, 1st or 2nd order transition $S$ $\rightarrow$ $P$, the order depending on g}
\item[$-$]{$T_{PF}$, Curie temperature of 2nd order transition $F$ $\rightarrow$ $P$}
\item[$-$]{$T_{SD}$, 2nd order transition $D$ $\rightarrow$ $S$}
\item[$-$]{$T_{FD}$, 2nd order transition $D$ $\rightarrow$ $F$}
\item[$-$]{$T_{1}$, $T^{*}$ are the end-points of the interval where $\Delta(T)$ is double-valued}
\end{itemize}
\begin{figure*}[h]
\subfigure[]{}\includegraphics[height=5.5cm]{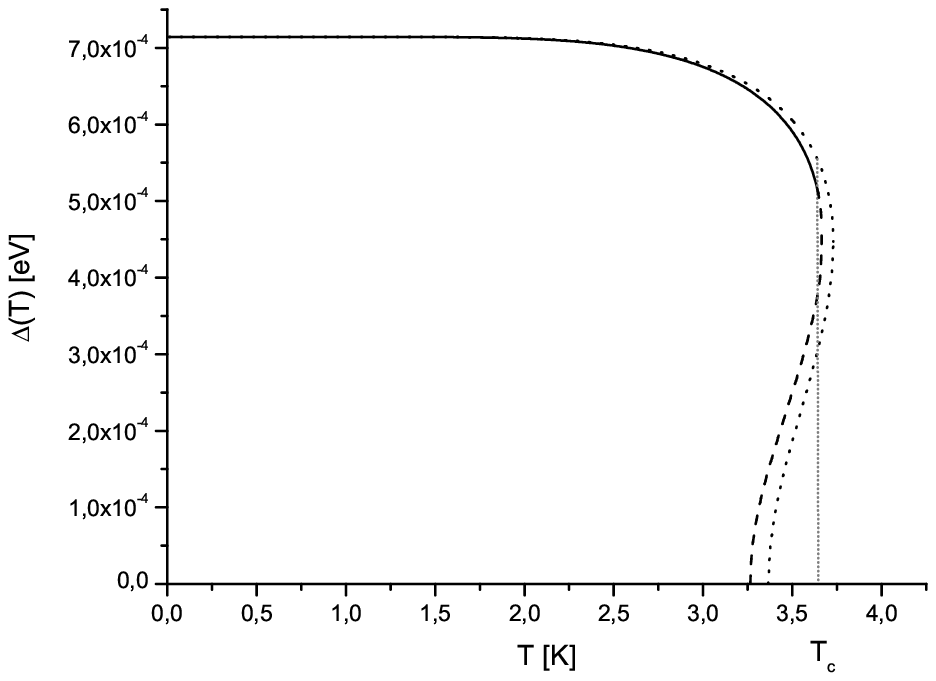}
\subfigure[]{}\includegraphics[height=6cm]{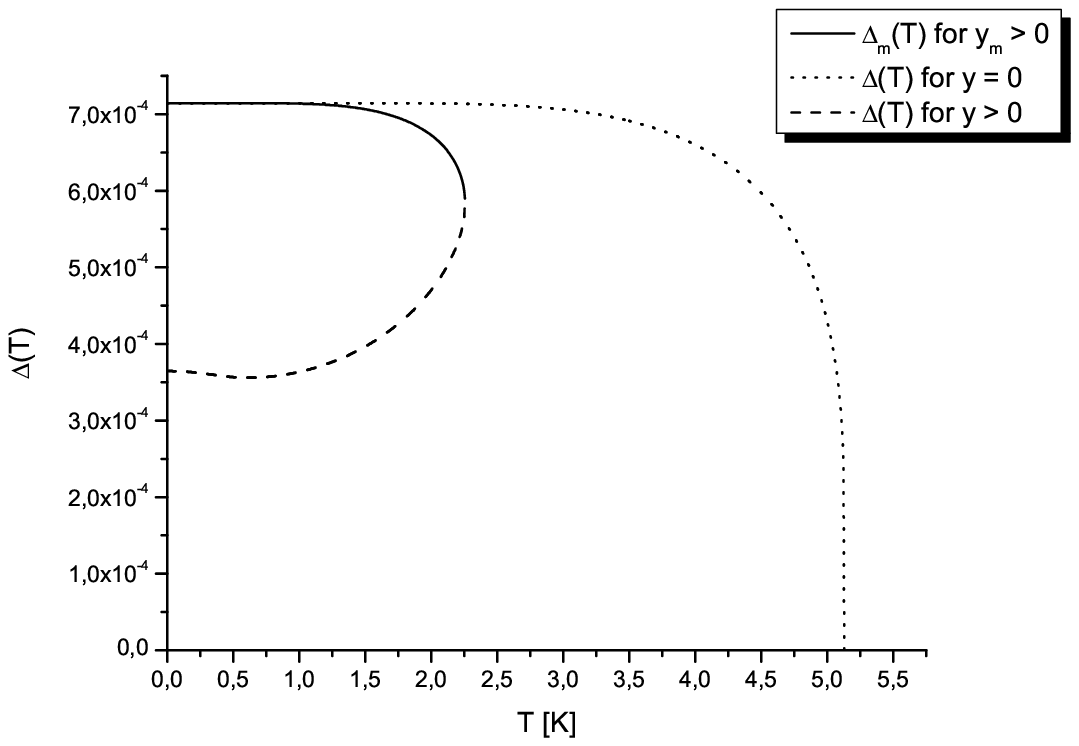}
\caption{\label{Delty}The gap parameter plotted as a function of reduced temperature $T$ for $G \varrho = 0.3, \delta = 0.01 \, eV, d = 0.05$ and: (a) $g = 0.01 \, \sqrt{\text{eV}}$, (b) $g = 0.03 \, \sqrt{\text{eV}}$. For sufficiently large value of $g$ the phase transition $S \rightarrow P$ is of the second order and $T_{\text{c}} = T^{*}$.}
\end{figure*}

The set of Eqs. (\ref{Delta}), (\ref{y}) has been solved numerically for different values of $g$ and impurity density $d$. FIG.~\ref{Delty} shows how variation of $g$ affects $\Delta(T)$. For solution with $\Delta \neq 0, y = 0$, $|T^{*} - T_{\text{c}}|$ declines as $g$ is increased and finally vanishes.  The temperature range, where $\Delta(T)$ is nonvanishing declines for the solution $\{\Delta \neq 0, y \neq 0\}$ with increasing value of $g$ and is smaller than for the solution $\{\Delta \neq 0, y = 0\}$. 

\begin{figure*}
\includegraphics{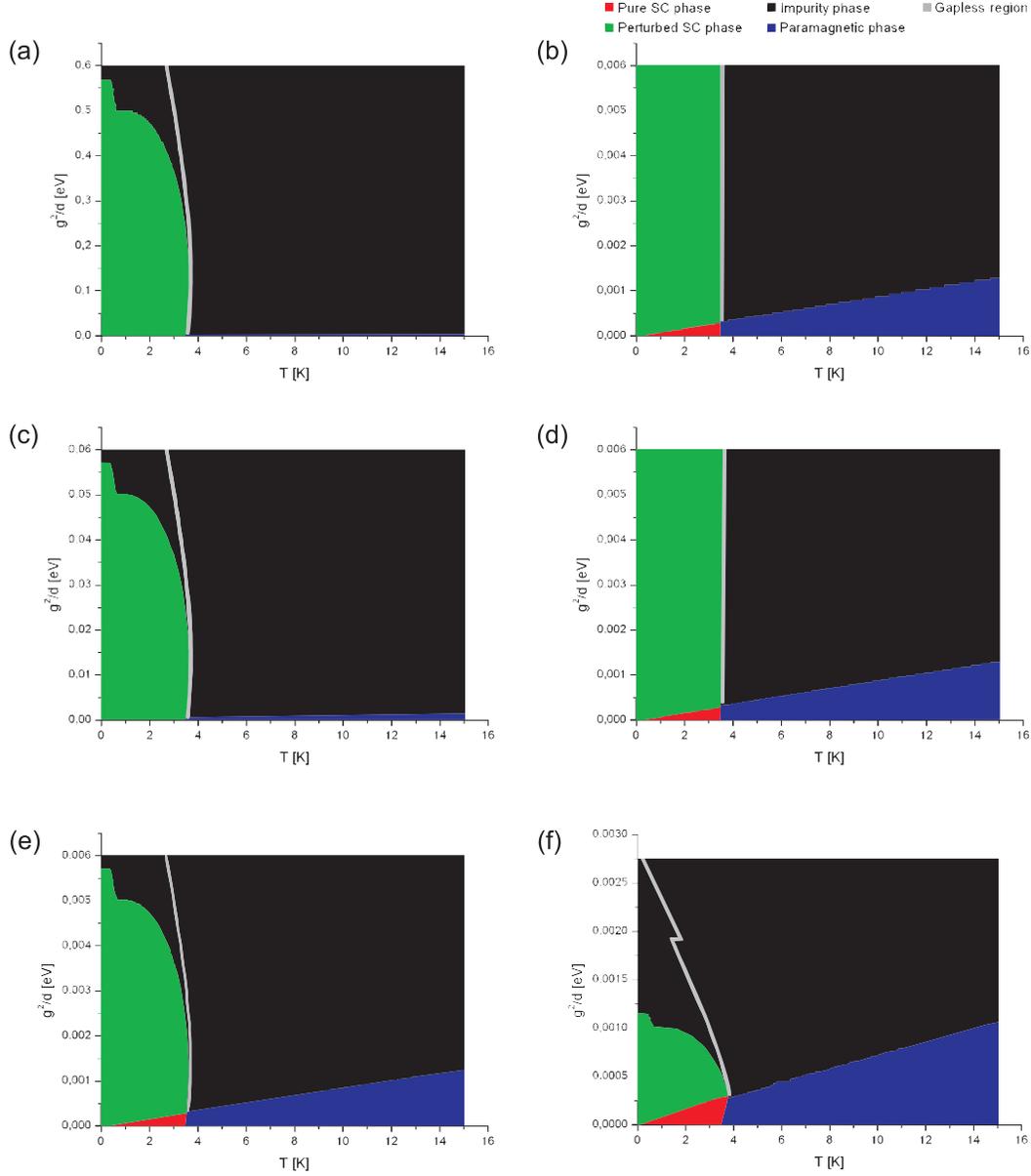}
\caption{\label{PD}Phase diagrams of the system for the following values of parameters: $G \varrho = 0.3, \, \delta = 0.01 \, \text{eV}$ and: (a) $d=0.001$, (b) $d=0.001$ (detailed version of (a) for small values of $g^{2}/d$) , (c) $d=0.01$, (d) $d=0.01$ (detailed version of (c)), (e) $d=0.1$, (f) $d=0.5$. The gapless region denotes the zone, where the smallest excitation energies from ground state disappear.}
\end{figure*}

The phase diagrams for fixed values of $G \varrho = 0.3, \, \delta = 0.01 \, \text{eV}$ are depicted in FIG.~\ref{PD}. These diagrams show the decline of the $D$ phase with increasing impurity concentration $d$. The $S$ phase and the $T_{\text{c}}$ temperature, on the other hand prove to be almost insensitive to variations of $d$, but one notices increase of $T_{\text{c}}$ with $g^{2}N^{-1} \propto g^{2}d^{-1}$ at higher $d$. In this respect the system's behaviour is rather unusual and may be explained as a consequence of the weakness of $V$ and breakdown of Eq.~(\ref{f}) at larger values of $d$. However, it is worth noting at this point that solid solutions of magnetic metals in zirconium and titanium have higher superconducting transition temperatures than pure zirconium and titanium \cite{MatthiasSuhl}. 

A subregion $NG$ with gapless superconductivity and $\Delta_{\text{m}} > 0$ (the light gray strip in FIG.~\ref{PD}) is present for sufficiently large values of $g^{2}d^{-1}$. The appearance of NG subregion is due to the negative terms $-g y$, $-2 g y$, which contribute to the spectrum of $h_{\underline{k}}$.

The perturbative effect of impurities increases with $g^{2}d^{-1}$ and causes the complete suppression of superconductivity at low temperatures, when $g^{2}d^{-1}$ is sufficiently large. At sufficiently low temperatures the mixed phase is present between the $S$ and $F$ phase, where superconductivity coexists with ferromagnetism of impurities. 

The temperature $T_{PF}$ decreases with impurity concentration $d$ and depends almost linearly on $g^{2}d^{-1}$. One also observes decrease of $T_{FD}$ with $g^{2}d^{-1}$ and increase of $T_{SD}$ with $g^{2}d^{-1}$. The temperatures $T_{SD}$, $T_{\text{c}}$, $T_{FD}$, $T_{PF}$ satisfy the following inequalities:
\[
T_{SD} \leq T_{\text{c}} \leq T_{PF} \quad \text{for small values of }g^{2}d^{-1}
\]
\[
T_{FD} \leq T_{PF} \quad \text{for large values of }g^{2}d^{-1}
\]
\section{Specific heat under varying impurity concentration}\label{SCV}
In terms of the energy density $u = \frac{\partial \beta f}{\partial \beta}$ the system's specific heat expresses as 
\begin{equation}
\label{CieploWlasciwe}
c = \frac{\partial u}{\partial T} + \frac{\partial u}{\partial y} \frac{\partial y}{\partial T} + \frac{\partial u}{\partial \Delta} \frac{\partial \Delta}{\partial T}.
\end{equation}
From Eq. (\ref{Gestoscf}) one obtains
\begin{equation}
\begin{split}
\label{EnergiaWewnetrzna}
u \left(\beta, E_{g}^{}, \xi, y \right) & = \varrho \int \limits_{-\delta}^{\delta} \Bigl[\tfrac{1}{2} \Delta^{2} E_{G}^{-1} \Bigl( F_{1}^{}\left(\beta, E_{G}^{}, \xi, y \right) + \beta \frac{\partial F_{1}}{\partial \beta}\Bigr)\\ &+ \tfrac{1}{2} (y^{2}+g^{2}) - \frac{1}{2}\frac{\partial \ln P}{\partial \beta} - g y \tanh (\beta g y d^{-1}) \Bigr]\text{d}\xi.
\end{split}
\end{equation}
The specific heat corresponding to a definite phase is denoted by the appropriate index, e.g. $c_{P}$ denotes the specific heat corresponding to paramagnetic phase $P$ ($\{\Delta_{\text{m}} = 0, y_{\text{m}} = 0 \}$).

\begin{figure}[h]
\subfigure[]{}\includegraphics[height=5.5cm]{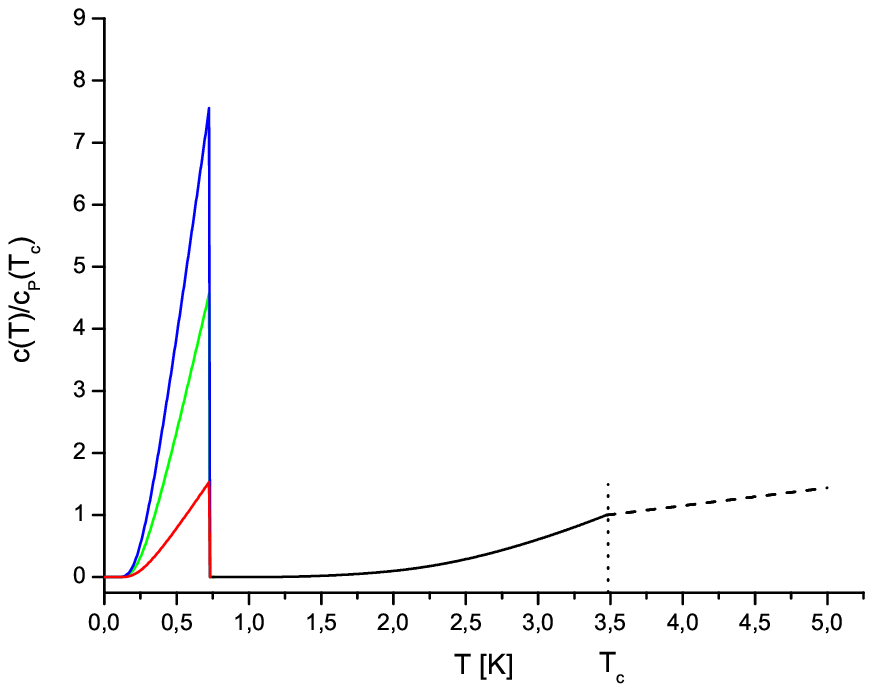}
\subfigure[]{}\includegraphics[height=5.7cm]{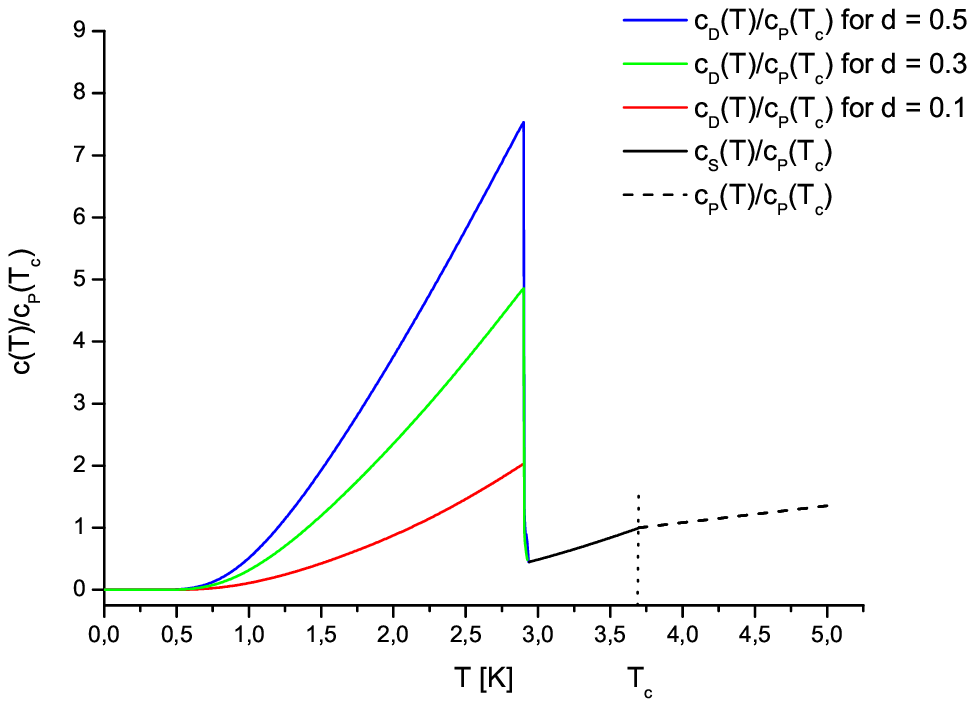}
\caption{\label{CV}The specific heat for different values of impurity concentration. ($G \varrho = 0.3, \, \delta = 0.01 \, \text{eV}$ and (a) $g^{2}d^{-1} = 6.25 \times 10^{-5} \, \text{eV}$, (b) $g^{2}d^{-1} = 2.50 \times 10^{-4} \, \text{eV}$)}
\end{figure}
The specific heat ratios $c(T)/c_{P}(T_c)$ under varying impurity concentration $d$ are depicted in FIG.~\ref{CV}. Similarly as in Ref.~\onlinecite{TM} no specific heat jump $\Delta c_{SP} = c_{S}-c_{P}$ at $T_{\text{c}}$ for sufficiently small values of magnetic coupling constant $g$ is observed and very weak dependence of $c_{S}$ and $c_{P}$ on $d$ has been found. FIG.~\ref{CV} also shows how the $S$ phase is replaced by the $D$ phase at low temperatures as $d$ increases and how the jump of $c$ at $T_{SD}$ increases with $d$.

\section{Chemical potential}\label{CP}
The properties of a superconductor can be determined from the solution of the gap equation, which is supplemented by the following equation for the chemical potential $\mu$
\begin{equation}
\label{mu}
\sum_{\underline{k} \sigma} \text{Tr}(n_{\underline{k}\sigma}^{} \rho_{0}^{}) = n,
\end{equation}
$n$ denoting the average number of fermions in the system and $\rho_{0}$ is the trial density matrix.
For a superconductor with 4-fermion attraction perturbed by magnetic impurities $\rho_{0}$ equals
\[
\rho_{0}^{} = \frac{\exp(-\beta h_{0}^{})}{\text{Tr} \exp(-\beta h_{0}^{})},
\]
and Eq.~(\ref{mu}) takes the form
\begin{equation}
\begin{split}
\sum_{\underline{k}\sigma} \text{Tr}(n_{\underline{k}\sigma}^{} \rho_{0}^{}) &= \tfrac{1}{2} \sum_{\underline{k} > 0} \text{Tr}\left[(n_{\underline{k}+}^{} +  n_{-\underline{k}+}^{} + n_{\underline{k}-}^{} + n_{-\underline{k}-})^{}\rho_{0}^{}\right] = -\tfrac{1}{4}\beta^{-1}\sum_{\underline{k}>0}\frac{\partial Z_{\underline{k}}}{ \partial \xi_{\underline{k}}} = \\ &= \tfrac{1}{2}\sum_{\underline{k}>0}\left[1 - \frac{\xi}{E_{G\underline{k}}} F_{1}(\beta, E_{G\underline{k}}^{}, \xi_{\underline{k}}^{}, y) - 2 F_{3}(\beta, E_{G\underline{k}}^{}, \xi_{\underline{k}}^{}, y)\right] = n.
\end{split}
\end{equation}
Thus, in the thermodynamic limit
\begin{equation}
\label{mu4f}
\int\limits_{-\mu}^{\infty} \varrho (\xi) \left[1 - \frac{\xi}{E_{G}}F_{1}^{}(\beta, E_{G}^{}, \xi, y) - 2 F_{3}^{}(\beta, E_{G}^{}, \xi, y) \right] \text{d}\xi = d,
\end{equation}
where
\begin{equation}
F_{3}^{}(\beta, E_{G}^{}, \xi_{\underline{k}}^{}, y) = \frac{\text{e}^{-\beta g^{2}/2} \sinh(\beta \xi) \cosh(\beta g y)}{4 \text{e}^{-\beta g^{2}/2} \cosh (\beta g y) \cosh (\beta \xi_{\underline{k}}^{}) + \text{e}^{-\beta g^{2}} \left[1 + \cosh (2 \beta g y) \right] + \cosh (2 \beta E_{G\underline{k}}^{}) + 1},
\end{equation}
and $d$ denotes the density of fermions in the free electron gas model, viz., $d = \tfrac{4}{3} \varepsilon_{F}^{} \varrho_{F}^{},\, \varrho_{F}^{} = m p_{F}(2 \pi^{2} \hbar^{2})^{-1}$.

The equation for $\mu$ at $T = 0$ results from Eq. (\ref{mu4f}) in the limit $T \rightarrow 0$, viz.,
\begin{equation}
\label{mu0}
\int \limits_{-\mu(0)}^{\infty} \varrho(\xi)\left[1 - \frac{\xi}{E_{G}} \right] \text{d}\xi = d.
\end{equation}
Substituting $E_{G}^{} = (\xi^{2} + \Delta^{2}(0))^{1/2}$, $\varrho(\xi) = \varrho_{F}^{} \chi(\xi)$ for $\xi \in [-\delta, \delta]$ and exploiting the equality $\varrho(\xi) = \varrho_{F}^{} \sqrt{\xi + \mu(0)}/\sqrt{\varepsilon_{F}}$ one obtains the equation for $\mu(0)$ [cf. Ref.~\onlinecite{TM}]
\begin{equation}
\label{mu02}
\tfrac{4}{3} \varepsilon_{F}^{-1/2} (\mu(0) - \delta)^{3/2} + 2 \delta = \tfrac{4}{3} \varepsilon_{F}^{}.
\end{equation}
For $\delta = 0.01$~eV, $\varepsilon_{F} = 1$~eV, Eq.~(\ref{mu02}) has the solution $\mu(0) = 0.9999748$~eV. For $\delta = 0.01$~eV, $\varepsilon_{F} = 1$~eV, $g = 0.01\,\sqrt{\text{eV}}$, $d = 
0.05,\,\Delta(T_{1})=y(T_{1})=0,$ Eq.~(\ref{mu4f}) has the solution $\mu(T_{1}) = 0.9999754$ eV, which is the same as $\mu(T^{*})$ for $\Delta(T^{*}) \neq 0$, $y(T^{*}) \neq 0$. It follows that the relative difference $|\mu(0) - \mu(T_{1})|/\mu(T_{1})$ is of order~$10^{-7}$ and  $|\mu(T_{1})-\varepsilon_{F}|$ is of order $10^{-5}$ eV. The very weak variation of $\mu(T)$ does not affect the temperature variation of $\Delta(T),\,y$, free energy and specific heat within the accuracy applied in numerical calculations.

\section{Conclusions}
We have shown that the thermodynamics of a superconductor with a quartet binding potential, perturbed by a reduced s-d interaction is solvable if the impurity density is sufficiently small. The essential properties of a superconductor perturbed by magnetic impurities have been demonstrated: decline of the temperature range, where the gap parameter is nonvanishing, with increasing coupling between conduction fermions and impurities, strong dependence of specific heat jump on the density of impurities at the transition to the $D$ phase with coexisting superconductivity and ferromagnetism and existence of gapless superconductivity. Due to the assumed weakness of the s-d coupling the value of $\Delta_{\text{m}}(0)$ and $T_{\text{c}}$ are not diminished by the impurities. These investigations will be extended to include the effect of a general s-d exchange interaction $V_{\text{s-d}}$.
\bibliography{4fermion}
\end{document}